# Lead Iodide Perovskite Light-Emitting Field-Effect Transistor

Xin Yu Chin,[1] Daniele Cortecchia,[2,3] Jun Yin,[1,4] Annalisa Bruno,[1,3]

and Cesare Soci[1,4,*]

[1] *Division of Physics and Applied Physics, School of Physical and Mathematical Sciences, Nanyang Technological University, 21 Nanyang Link, Singapore 637371*

[2] *Interdisciplinary Graduate School, Nanyang Technological University, Singapore 639798*

[3] *Energy Research Institute @ NTU (ERI@N), Research Technoplaza, Nanyang Technological University, 50 Nanyang Drive, Singapore 637553*

[4] *Centre for Disruptive Photonic Technologies, Nanyang Technological University, Nanyang, 21 Nanyang Link, Singapore 637371*

*Corresponding author: csoci@ntu.edu.sg

**Abstract**

Despite the widespread use of solution-processable hybrid organic-inorganic perovskites in photovoltaic and light-emitting applications, determination of their intrinsic charge transport parameters has been elusive due to the variability of film preparation and history-dependent device performance. Here we show that screening effects associated to ionic transport can be effectively eliminated by lowering the operating temperature of methylammonium lead iodide perovskite ($CH_3NH_3PbI_3$) field-effect transistors (FETs). Field-effect carrier mobility is found to increase by almost two orders of magnitude below 200 K, consistent with phonon scattering limited transport. Under balanced ambipolar carrier injection, gate-dependent electroluminescence is also observed from the transistor channel, with spectra revealing the tetragonal to orthorhombic phase transition. This first demonstration of $CH_3NH_3PbI_3$ light-emitting FETs provides intrinsic transport parameters to guide materials and solar cell optimization, and will drive the development of new electro-optic device concepts, such as gated light emitting diodes and lasers operating at room temperature.

**Introduction**

Organolead halide perovskites are emerging solution-processable materials with outstanding optoelectronic properties.[1-7] Among them, methylammonium lead iodide $CH_3NH_3PbI_3$ has proven to be an exceptional light harvester for hybrid organic-inorganic solar cells,[3, 8-15] which in just four years achieved an impressive NREL-certified power conversion efficiency of 20.1%, and remarkable performance in a variety of device architectures.[16] Thanks to their cost-effectiveness and ease of processing, hybrid perovskites have naturally attracted a vast interest for applications beyond photovoltaic energy conversion, such as water splitting,[17] light-emitting diodes[18-20] and tunable, electrically pumped lasers.[6, 21-23] So far transport parameters of perovskite materials were mostly deduced from the study of photovoltaic devices, which indicated ambipolar transport[3, 24, 25] of holes and electrons within the perovskite active region, and long electron-hole pair diffusion length.[4, 5, 26] First-principle calculations for this class of materials predict that hole mobility is up to 3100 $cm^2 V^{-1} s^{-1}$ and electron mobility is 1500 $cm^2 V^{-1} s^{-1}$ with concentration of $10^{16}$ $cm^{-3}$ at 400 K,[50] and high frequency mobility of 8 $cm^2 V^{-1} s^{-1}$ was determined in $CH_3NH_3PbI_3$ spin coated thin film by THz spectroscopy,[27] a remarkably high value for solution-processed materials. A combination of resistivity and Hall measurement further revealed that the mobility of ~66 $cm^2 V^{-1} s^{-1}$ are achievable in $CH_3NH_3PbI_3$.[28]

However, very recently ion drift was shown to play a dominant role on charge transport properties,[29] stimulating an ongoing debate about the carrier character and the origin of anomalous hysteresis, together with the role of polarization, ferroelectric, and trap-state filling effects in organolead halide perovskite devices investigated at room temperature.[30-33]

Despite the rapid advancement of optoelectronic applications, a big gap remains in understanding the fundamental transport properties of organolead halide perovskites, namely charge carrier character, mobility and charge transport mechanisms. To fill this gap, studies of basic field-effect transistor (FET) devices are urgently needed. Historically, related tin(II) based 2D hybrid perovskites have attracted major interest for FET fabrication due to their attractive layered structure, with demonstrated field-effect mobilities up to 0.62 cm$^2$ V$^{-1}$ s$^{-1}$ and $I_{on}/I_{off}$ ratio above 10$^4$.[34] Improvement of mobility can be achieved by substitution of organic cation in hybrid perovskite, yielding FET saturation-regime mobility as high as 1.4 cm$^2$ V$^{-1}$ s$^{-1}$, and nearly an order of magnitude lower linear-regime mobility.[35] Further improvement was demonstrated through melt processed deposition technique, where saturation and linear mobilities of 2.6 and 1.7 cm$^2$ V$^{-1}$ s$^{-1}$ with $I_{on} / I_{off}$ of 10$^6$ was achieved.[36] Conversely, only rare examples of 3D hybrid perovskites FETs can be found in the literature,[15] with limited hole mobility of the order of ~10$^{-5}$ cm$^2$ V$^{-1}$ s$^{-1}$ in the case of CH$_3$NH$_3$PbI$_3$ and strong hysteresis due to ionic transport, which so far have hindered the development of FET applications. Nonetheless, the high photoluminescence efficiency[22] and widely tunable band gap from visible to infrared[28,37] make CH$_3$NH$_3$PbI$_3$ extremely attractive for the fabrication of solution processable light-emitting field-effect transistors (LE-FET), a device concept that may be integrated in heterogeneous optolectronic systems, such as flexible electroluminescent displays[38] or electrically pumped lasers.[39]

Here we report the fabrication and characterization of CH$_3$NH$_3$PbI$_3$ field-effect transistors, and their operation as light-emitting FETs yielding efficient gate-assisted electroluminescence. Low-temperature measurements were used to effectively remove screening effects arising from ionic transport, allowing the determination of intrinsic transport parameters such as carrier density and mobility. Field-effect mobility of $CH_3NH_3PbI_3$ is found to increase by almost 2 orders of magnitude from room temperature down to 78 K, a behavior consistent with phonon scattering limited transport of conventional inorganic semiconductors. We also confirm the ambipolar nature of charge transport in $CH_3NH_3PbI_3$, which yields close to ideal ambipolar transistor characteristics and electroluminescence from the transistor channel under balanced injection conditions. To the best of our knowledge, this is the first demonstration of $CH_3NH_3PbI_3$ light-emitting FETs. In addition to providing an essential guideline for materials optimization through chemical synthesis and future improvements of solar cell performance, this novel device concept opens up new opportunities for the development of electro-optic devices based on $CH_3NH_3PbI_3$, such as gated, electrical injection light-emitting diodes and lasers operating at room temperature.

Deposition methods of solution-processed organo-lead hybrid perovskite have direct consequences on the morphology of thin film, hence the charge transport properties of the material.[2] Here we used the solvent engineering technique recently reported for optimized solar cell fabrication[14] to deposit a compact and uniform $CH_3NH_3PbI_3$ perovskite layer (~150 nm thick) on top of heavily p-doped Si with thermally grown $SiO_2$ (**Figure 1a**). The resulting thin films are of very high quality: they consist of closely-packed, large domains with grain size up to 200 nm, as seen in the top view SEM image in **Figure 1b**, which crystallize in a perfect tetragonal structure, as revealed by the XRD analysis in **Figure 1c**. Availability of such high quality films is essential to minimize the influence of metal contacts and charge carrier scattering across the film, so as to obtain intrinsic transport parameters from FET measurements. The device structure used in this study shown in **Figure 1d**. A bottom gate, bottom contact configuration was employed to allow deposition of active materials to be the last step in the fabrication. This is to minimize exposure of $CH_3NH_3PbI_3$ to moisture in the environment, and to avoid potential overheating during the metal electrode deposition.

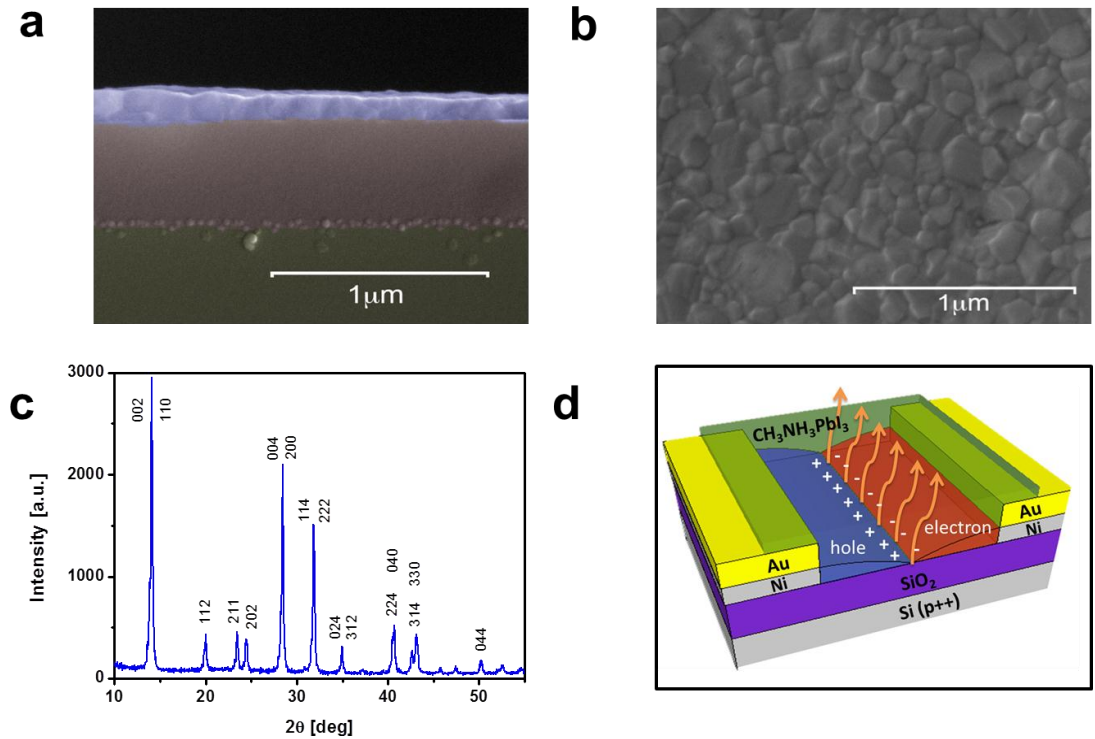

**Figure 1 | FET device configuration and thin film characterization. a,b** Cross sectional (**a**) and top-view (**b**) SEM micrographs of the $CH_3NH_3PbI_3$ thin film. **c,** XRD pattern of $CH_3NH_3PbI_3$ film on $SiO_2$/Si(p++) substrate, confirming the tetragonal structure of the perovskite and space group *I*4*/mcm*. **d**, Schematic of the bottom-gate, bottom contact LE-FET configuration used in this study.

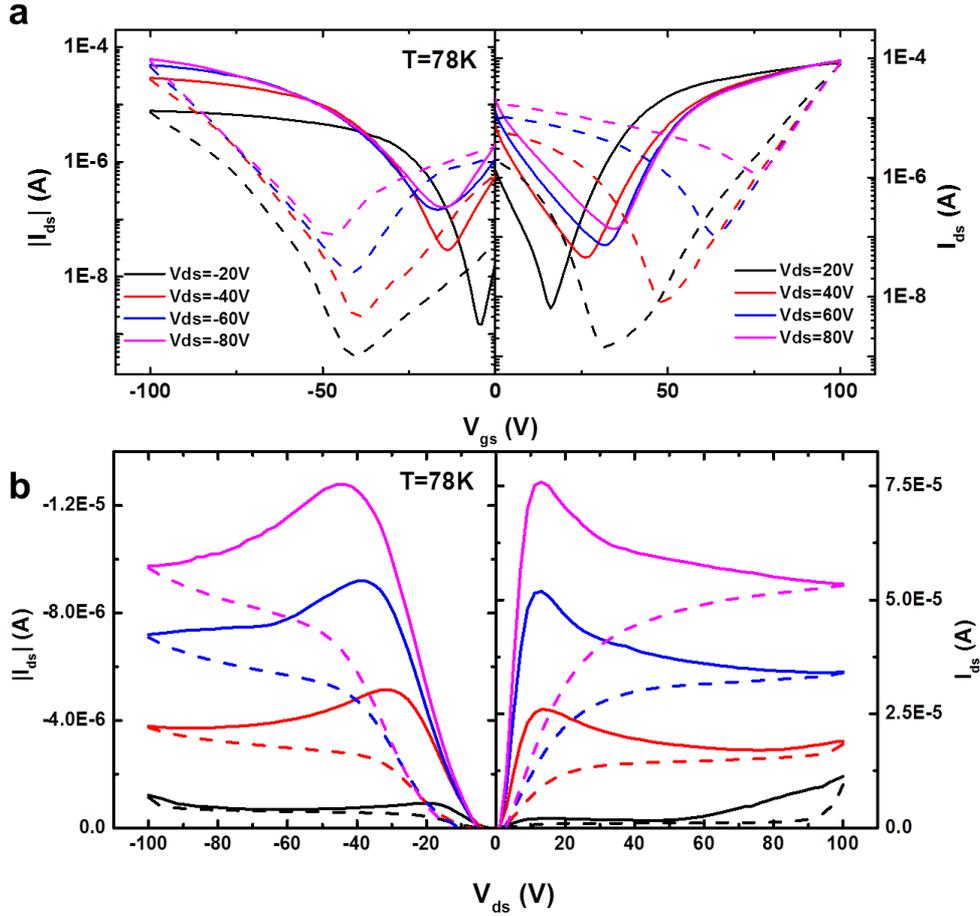

**Figure 2 | FET characteristics. a,b** Transfer (**a**) and output (**b**) characteristics obtained at 78 K. The n-type output characteristics (right panel) were measured at $V_{gs}$=40 V to 100 V ($V_{gs}$=40 V black, $V_{gs}$=60 V red, $V_{gs}$=80 V blue, $V_{gs}$=100 V magenta), while the p-type output characteristics (left panel) are measured at $V_{gs}$= - 40 V to - 100 V ($V_{gs}$= - 40 V black, $V_{gs}$= - 60 V red, $V_{gs}$= - 80 V blue, $V_{gs}$= - 100 V magenta). Solid and dashed curves are measured with forward and backward sweeping, respectively. See supplementary information for the full set of FET characteristics as a function of temperature.

As reported in the literature, transport characteristics of $CH_3NH_3PbI_3$ solar cells are subject to strong hysteresis, which so far hindered a complete understanding of the electrical response, and the determination of intrinsic transport parameters of the perovskite.[30-32] The origin of this anomalous behavior has been attributed to capacitive effects associated with ferroelectricity arisen from the spontaneous polarization of methylammonium cation and lattice distortion effects, diffusion of excess ions as interstitial defects, and trapping/de-trapping of charge carriers at the interface.[30, 31, 32] Recently, photocurrent hysteresis in $CH_3NH_3PbI_3$ planar heterojunction solar cells was found to be originated from trap states on the surface and grain boundaries of the perovskite materials, which can be effectively eliminated by fullerene passivation.[32] Piezoelectric microscopy revealed the reversible switching of the ferroelectric domains by poling with DC biases,[40] but a recent observation of field-switchable photovoltaic effect suggested that ion drift under the electric field in the perovskite layer induces the formation of p–i–n structures,[29] as observed by electron beam-induced current measurement (EBIC) and Kelvin probe force microscopy (KPFM).[24, 25] A weakened switchable photovoltaic effect at low temperature and the lack of photovoltage dependence with respect to the lateral electrode spacing suggest that ferroelectric photovoltaic effect may not play dominant role in the observed field-switchable photovoltaic behavior.[29] Theoretical calculations further reveal that charged Pb, I, and methylammonium vacancies have low formation energies[40, 41], suggesting that the high ionicity of this materials may lead to p- and n-type self-doping.

We found that reducing the operating temperature of our devices is an effective way to reduce hysteresis effects due to ionic transport/screening, allowing to record transport characteristics typical of conventional ambipolar semiconductor FETs (**Figure 2**). The complete temperature evolution of ambipolar FET characteristics, from room temperature down to 78 K, is provided in **Figures S1** and **S2** of the supplementary information. While above 198 K the output characteristics show either weak or no gate voltage dependence, at and below 198 K the devices display "textbook" n-type output characteristics. Similarly, typical p-type output characteristics are observed at 98 K and lower temperatures (**Figures 2a** and **S2**). Both p- and n-type transfer characteristics are independent of gate field from room temperature down to 258 K. This is reflected in the measurement by large hysteresis loops, which do not close when transitioning from the hole- to the electron-dominated transport gate voltage ranges and vice versa. Below 258 K, however, both n- and p-type transfer characteristics show a closed hysteresis loop. Hysteresis of n- and p-type transfer characteristics is substantially reduced below 198 K and 98 K, respectively, consistent with the observation of ambipolar output characteristics (**Figures 2b** and **S2**). Induced carrier density of ~$3.8\times10^{16}$ cm$^{-2}$, maximum $I_{on}$ / $I_{off}$ ~ $10^5$, and current density of ~ 830 A cm$^{-2}$ (estimated for a ~2 nm accumulation layer thickness) are obtained from standard transistor analysis at 198 K. These values are comparable to those previously reported for 2D hybrid organic-inorganic perovskites characterized at room temperature.[35, 36] Note that, although our low-temperature measurements clearly demonstrate the ambipolar nature of CH$_3$NH$_3$PbI$_3$, previous studies have shown that carrier concentration can vary by up to six orders of magnitudes depending on the ratio of the methylammonium halide and lead iodine precursors and thermal annealing conditions, thus resulting in preferential p-type or n-type transport characteristics.[41]

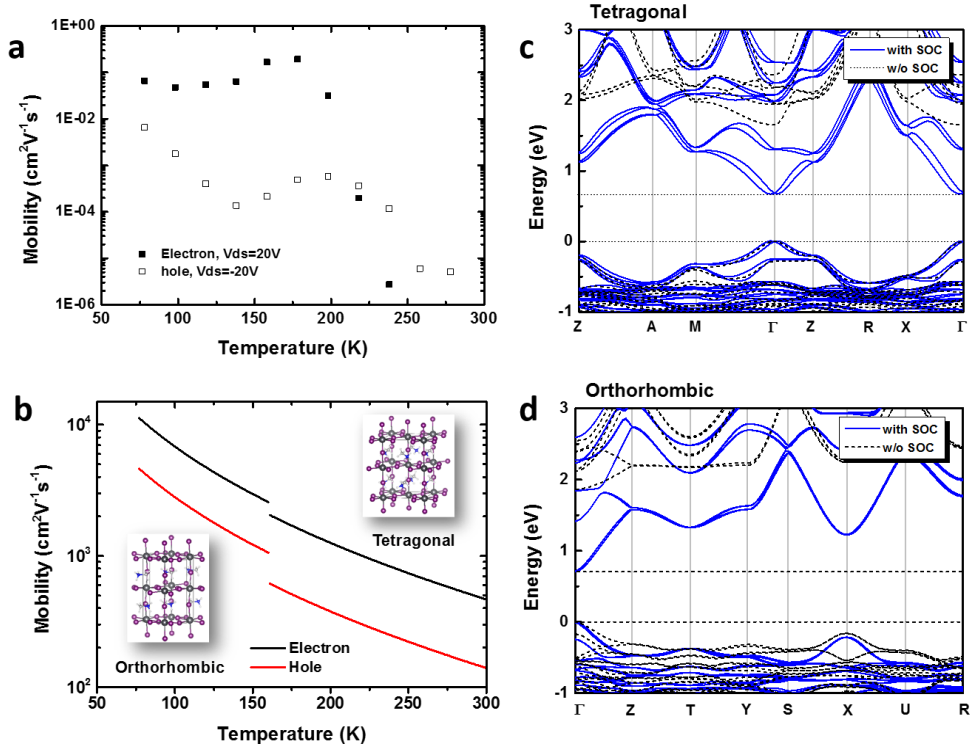

**Figure 3 | Experimental and theoretical field-effect mobility and band structures of CH$_3$NH$_3$PbI$_3$.**
**a**, Temperature dependence of field-effect electron and hole mobilities, extracted from the forward sweeping of transfer characteristics at $V_{ds}$ = 20 V and $V_{ds}$ = - 20 V, respectively. **b,** Calculated temperature dependence hole (red curves) and electron (black curves) mobility in tetragonal (T=300-160 K) and orthorhombic (T=160-77 K) phases of CH$_3$NH$_3$PbI$_3$. The crystal unit cells of the two phases are shown as insets. **c, d,** Band structures of the tetragonal (**c**) and orthorhombic (**d**) phases obtained by DFT-PBE method with (solid curves) and without (dotted curves) spin-orbital coupling (SOC).

Temperature dependent electron and hole mobilities were extracted from the forward sweeping of transfer characteristics at $V_{ds}$ = 20 V and $V_{ds}$ = - 20 V using the standard transistor equation at linear regime.[42] The resulting values are shown in **Figure 3a**.[#] A statistical analysis of the distribution of mobility values extracted from independent measurements across 4 different devices is also available in supplementary **Figure S3**. While some variability in the absolute values of electron and hole mobilities is observed from device to device, their relative magnitude and temperature dependence show consistent trends. From **Figure 3a**, both electron and hole mobilities increase by a factor of ~100 from room temperature to 198 K. Below 198 K there is no further improvement of electron mobility, while hole mobility shows an additional tenfold increase. We attribute the improvement of mobility at low temperature to the removal of screening effects arising from the ionic transport of methylammonium cations. The phonon energy of methylammonium cation was estimated to be ~14.7 meV from previous combination of DFT and Raman studies.[43] The observation of weak improvement of field-effect mobilities below 198 K ($E_{thermal}$=16.7 meV) is therefore consistent with the quenching of phonon interactions related to the organic cations. This is also in agreement with the weakening of field-switchable photovoltaic effects at low temperature,[29] strongly suggesting that field-effect transport is phonon- limited at room temperature. Despite the remarkable improvement of field-effect mobilities, hysteresis was not completely removed at the lowest temperature investigated. This could be due to the untreated semiconductor–dielectric interface, which is known to affect semiconductor film morphology, number of trap states, and surface dipoles, similar to the case of organic field-effect transistor devices.[42] Further investigations will be required to address this issue. Both hole and electron mobilities extracted in the linear regime at 78 K are slightly smaller than the corresponding saturation regime mobilities ($\mu_{e,linear} / \mu_{e,saturation}$ = 6.7×10$^{-2}$ / 7.2×10$^{-2}$ cm$^2$ V$^{-1}$ s$^{-1}$ and $\mu_{h,linear} / \mu_{h,saturation}$= 6.6×10$^{-3}$ / 2.1×10$^{-2}$ cm$^2$ V$^{-1}$ s$^{-1}$, extracted at $V_{ds}$ = ± 20 V for linear regime and $V_{ds}$ = ± 80 V from saturation regime from **Figure 2a**). A previous study of spin-coated hybrid perovskite channels indicated linear regime mobility values 1 to 2 orders of magnitude lower than in the saturation regime.[35] The suppression of the linear regime mobility is presumably associated to grain-boundary effects, which give rise to a large concentration of traps. Thus, our reported linear regime mobilities set a lower limit for electron and hole mobilities of $CH_3NH_3PbI_3$.

---

[#] Note that mobilities were not extracted from backward sweeping curves to avoid misleading results due to the large hysteresis. Also, mobilities at higher $V_{ds}$ (*i.e.* in the saturation regime) were not extracted due to the difficulty to identify linear and saturation regimes at all investigated temperatures.

To better understand the transport data, we estimated the mobility of $CH_3NH_3PbI_3$ for both tetragonal and orthorhombic crystallographic phases using semi-classical Boltzmann transport theory,[44] upon deducing charge carrier effective masses and electron (hole)-phonon coupling. Electron and hole effective masses listed in **Table S1** were derived by quadratic fitting of the band structure dispersion (**Figures 3c** and **3d**); the corresponding fitting parameters are summarized in **Table S2**. The average effective mass of electrons (tetragonal: 0.197 $m_0$, orthorhombic: 0.239 $m_0$) is consistently smaller than the one of holes (tetragonal: 0.340 $m_0$, orthorhombic: 0.357 $m_0$). The resulting mobilities (**Figure 3b**) increase at lower temperatures due to the Boltzmann activation energy (see Computational Methods section), in good agreement with our experimental results. Although the calculated mobilities are substantially larger than the experimental values in **Figure 3a**, calculations reflect fairly well the relative magnitude of electron vs hole mobility, as well as the different mobility of the two crystallographic phases. Within the entire temperature range investigated, electron mobilities exceed hole mobilities by approximately a factor of two, and increase by nearly one order of magnitude below the phase transition temperature ($\mu_e$=2577−11249 cm$^{-2}$ V$^{-1}$ s$^{-1}$ and $\mu_h$=1060−4630 cm$^{-2}$ V$^{-1}$ s$^{-1}$ for the orthorhombic phase and $\mu_e$=466−2046 cm$^{-2}$ V$^{-1}$ s$^{-1}$ and $\mu_h$=140−614 cm$^{-2}$ V$^{-1}$ s$^{-1}$ for the tetragonal phase). The small experimental values can partly be attributed to the increase of effective masses by elastic carrier−phonon scattering, which is expected in real crystals due to defects and disorder induced by the organic components, as well as carrier-carrier scattering at high electron and hole concentrations.[45] Formation of segregation pathways for hole and electron transport due to the ferroelectric methylammonium cation could also elongate the carrier drifting path, hence lower carrier mobilities.[46] In addition, polycrystalline domains typical of solution-processed $CH_3NH_3PbI_3$ thin films (**Figures 1a** and **1b**) are likely to weaken the electronic coupling between grains, requiring charge carriers to hop along and across domain boundaries, further reducing the effective carrier mobility.

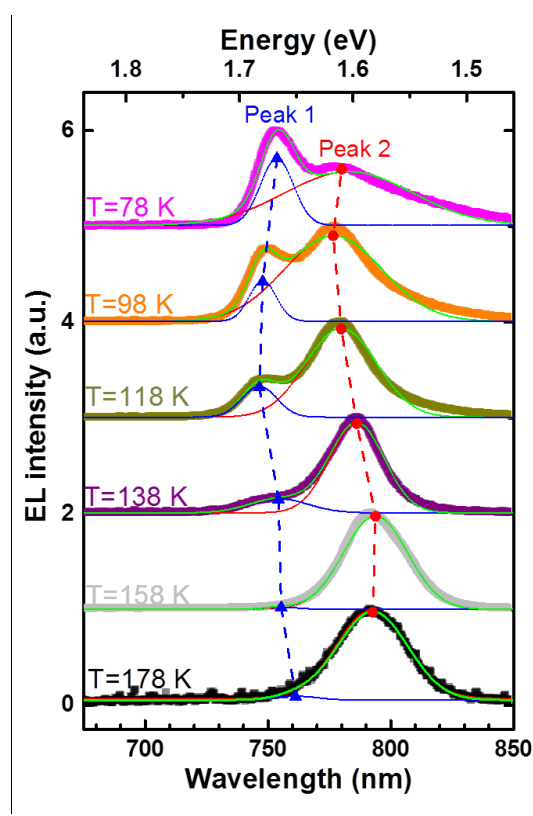

**Figure 4 | Low-temperature electroluminescence spectra of $CH_3NH_3PbI_3$ LE-FET.** EL spectra (collected at $V_{ds}$=100 V, $V_{gs}$=100 V) were normalized to their maximum peak. The spectra were fitted by two Gaussian curves (solid lines). The shift in peak position of the 750 nm peak (namely Peak 1, blue triangles) and 780 nm peak (Peak 2, red circles) is highlighted by connecting dashed lines.

The excellent ambipolar characteristics shown by the $CH_3NH_3PbI_3$ FET at low temperature (**Figure 2**) are rather encouraging for the realization of light emitting devices operating under balanced carrier injection.[6, 18-23] In particular, large carrier injection via charge accumulation at the semiconductor-dielectric interface is known to be an effective way to achieve bright and fast-switchable electroluminescence, and to optimize the spatial location of the carrier recombination zone in organic gate-assisted light-emitting field-effect transistors (LE-FETs).[47] In LE-FET devices, ambipolar channels are formed simultaneously by proper source-drain and gate biasing. Under perfectly balanced conditions, holes and electrons injected from opposite electrodes recombine in the middle of the FET channel, thus defining a very narrow radiative emission zone, as depicted in **Figure 1d.** The brightness of emission as well as the spatial position of the radiative recombination zone can be tuned by gate and drain-source biases.[42] LE-FET structures have proved to improve the lifetime and efficiency of light-emitting materials thanks to the large electrical injection achievable, and the possibility to optimize and balance charge carrier recombination compared to conventional LED devices.[38, 48] Combined with the ease of integration as nanoscale light sources in optoelectronic and photonic devices, this makes LE-FETs a very promising concept for applications in optical communication systems, solid-state lighting, and electrically pumped lasers.[48, 49]

Indeed, our $CH_3NH_3PbI_3$ FETs show substantial light emission when operated in their ambipolar regime at low temperature (78-178 K). Typical electroluminescence (EL) spectra are displayed in **Figure 4**. Note that no light emission could be observed above 198 K, most likely due to the large ionic screening effects discussed earlier, so that low temperature operation is necessary at this stage. The emission spectra of the LE-FET are consistent with direct recombination of injected electrons and holes into the perovskite active region. At the lowest temperature investigated (78 K) the EL spectrum shows two peaks centred at 750 nm (Peak 1) and 780 nm (peak 2), with distinct amplitudes and spectral positions at different temperatures. While Peak 1 appears only at temperatures below 158 K, Peak 2 dominates the EL spectra at higher temperatures. A similar behaviour was already reported for temperature dependent photoluminescence measurements of $CH_3NH_3PbI_3$,[50] and related to a structural transition from a low-temperature orthorhombic phase to a high-temperature tetragonal phase occurring around 162 K, as predicted by density functional theory.[51,52] Occurrence of this phase transition in the temperature regimes of 150–170 K for $CH_3NH_3PbI_3$ and 120–140 K for hybrid $CH_3NH_3PbI_{3-x}Cl_x$ was also confirmed by low temperature absorption studies.[6] Peak 1 and Peak 2 in our EL measurements are then assigned to the low-temperature orthorhombic phase, and to the high-temperature tetragonal phase, respectively. DFT calculations (**Figures 3c** and **3d**) reveal that the bandgap of the tetragonal phase is smaller than the orthorhombic phase of $CH_3NH_3PbI_3$, consistent with the energy of the EL peaks (Peak 1: 1.65 eV and Peak 2: 1.59 eV). The simultaneous presence of both Peak 1 and Peak 2 might indicate significant phase coexistence in the hybrid perovskite films, particularly at intermediate temperatures. To quantify the relative intensity and spectral energy of the two emission peaks as a function of temperature, we analysed the EL spectra by a deconvoluted Gaussian fitting (see Gaussian curves in **Figure 4** and corresponding fitted parameters in supplementary **Figure S4**). While both peaks show the expected red shift at the lowest temperatures, their temperature dependence in the intermediate range 118-178 K is rather complicated (**Figure S4a**). Moreover, while the Gaussian FWHM of Peak 1 reduces at lower temperatures, the FWHM of Peak 2 shows the opposite behavior (**Figure S4b**), as previously seen in low temperature photoluminescence measurements.[50] At this stage the anomalous spectral shift and broadening of the EL peaks as a function of temperature are not completely understood, and further investigations are needed to reveal their nature.

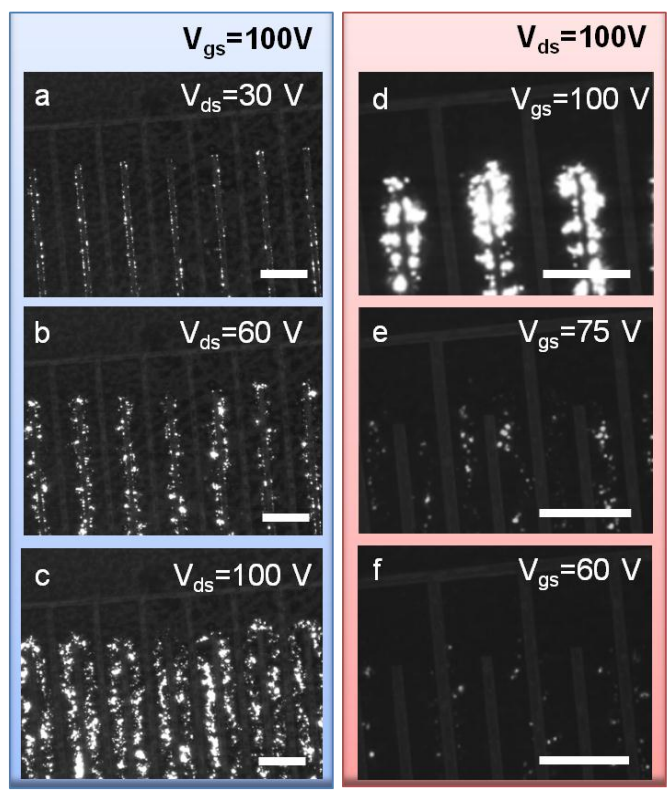

**Figure 5 | Optical images of CH$_3$NH$_3$PbI$_3$ LE-FET emission zone at T = 158 K. a,b,c** Frame images extracted from a video recorded while sweeping $V_{ds}$ from 0 to 100 V at constant $V_{gs}$=100 V; the corresponding values of $V_{ds}$ are indicated in the panels. **d,e,f** Frame images extracted from a video recorded while sweeping $V_{gs}$ from 0 to 100 V at constant $V_{ds}$ = 100 V; the corresponding values of $V_{gs}$ are indicated in the panels; note that the contrast of the metal contacts was slightly enhanced for clarity. See supplementary **Videos S1** and **S2** for the source real time videos of the measurements. Scale bars: 200 μm.

To achieve simultaneous hole and electron injection in a LE-FET, the local gate potential at drain and source electrodes must be larger than the threshold voltage of either of the charge carrier (*i.e.* $|V_d| > |V_{th,h}|$ and $V_s > V_{th,e}$, or $V_d > V_{th,e}$ and $|V_s| >$ $|V_{th,h}|$).[42] Under this condition, drain-source and gate voltages are tuned to control the injected current density of both carriers, which manipulate the spatial position of the emission zone as well as the EL intensity.[42] **Figure 5** shows microscope images of the emission zone of the LE-FET recorded at 158 K under different biasing conditions. Despite the grainy light emission pattern due to the polycrystalline nature of the film (**Figures 1a** and **1b**), the EL emission zone can be clearly identified from the images. For a fixed gate bias of $V_{gs}$ = 100 V (**Figures 5a** to **5c**), the emission zone is mainly concentrated near the drain electrode when $V_{ds}$ is small (**Figure 5a**). This is due to the limited injection of holes resulting from the relative low absolute local gate potential at the drain electrode $|V_d|$. By increasing $V_{ds}$, $|V_d|$ increases, thus more holes are injected into the active channel and the EL intensity increases (**Figure 5b**). Further increase of hole injection extends the emission area to the center of the channel, enhancing the EL intensity even further (**Figure 5c**). Conversely, for a fixed drain-source voltage of $V_{ds}$ = 100 V (**Figures 5d** to **5f**), the injected electron and hole current densities can no longer be regulated independently. **Figure 5d** shows extremely bright emission from nearby the drain electrodes due the overwhelming density of injected electrons recombining with a comparatively lower density of injected holes. Decreasing the gate voltage reduces the local gate potential at the source electrode $V_s$ and increases $|V_d|$, thus decreasing electron injection and increasing hole injection. This pushes the emission zone to the center of the active channel and reduces the EL intensity since overall current density decreases (**Figure 5e**). A further reduction of gate voltage pushes the emission zone closer to the source electrode, further weakening the EL intensity (see **Figure 5f**). Continuous-frame videos showing the variation of EL intensity and position of the emission zone sweeping $V_{ds}$ from 0 to 100 V at constant $V_{gs}$ = 100 V and sweeping $V_{gs}$ from 0 to 100

V at constant $V_{ds}$ = 100 V are provided as supplementary **Videos S1** and **S2**. This demonstrates that full control of charge carrier injection and recombination in $CH_3NH_3PbI_3$ LE-FET can be easily achieved by adjusting its biasing conditions, a necessary step toward the realization of hybrid perovskite electrical injection lasers.

In summary, we fabricated high-quality hybrid perovskite FETs and used them to determine intrinsic transport parameters of $CH_3NH_3PbI_3$, which are of great relevance to electro-optic devices (including solar cells). Our main findings include the ambipolar nature of charge transport, the understanding of the origin and suppression of screening effects associated to the presence of ionic cations, the direct determination of electron and hole mobilities and their temperature dependence, and the effect of structural phase transition on the electronic properties of $CH_3NH_3PbI_3$, all in good agreement with first-principle DFT calculations. Furthermore, bright electroluminescence due to radiative recombination within the transistor channel was demonstrated under balanced charge injection. We believe this first demonstration of a $CH_3NH_3PbI_3$ light-emitting field-effect transistor paves the way to the realization of solution-processed hybrid perovskite light emitting devices such as high-brightness light-emitting diodes and electrical injection lasers. More work will be needed in this direction to minimize ionic screening, improve thin film crystallinity and optimize device architecture, for instance employing staggered FET configurations to increase carrier injection density,[42] or integrating surface microstructures for light management.

**Materials and Methods**

*FET fabrication:* Heavily p-doped Si substrates with thermally grown $SiO_2$ (500 nm) layer were cleaned by two rounds of sonication in acetone and iso-propyl alcohol (20 minutes each round, and then dried under nitrogen flow. Interdigitated electrodes (L = 80 μm and 100 μm, W = 20 mm) were patterned using conventional photolithography. Electrodes of Ni (10 nm) and Au (50 nm) were thermally evaporated. The substrates were then undergoing lift-off process to obtain the desired electrodes. Before the spin coating of the active materials, an oxygen plasma cleaning treatment was performed on the substrate, for 1 minute, to improve the wetting of the surface and obtain flatter and homogeneous perovskite thin film. (See perovskite deposition)

*Temperature dependence FET measurements:* FET devices were mounted into a liquid nitrogen cooled Linkam Stage (FTIR 600) that allow to scan FET operating temperature of the device from 300 K down to 77 K. The FET electrical characteristics were acquired with Agilent B2902A Precision Source/Measure Unit in dark environment. The data were then analyzed with OriginPro software.

*Electroluminescence measurement:* The EL spectra were acquired using the Nikon eclipse LV100 microscope with LU plan fluor 10x objectives while the FET were enclosed in the Linkam Stage and FET electrical behavior was controlled using Agilent B2902A Precision Source/Measure Unit. EL emission signal was focused into optic fiber that coupled to USB2000 Ocean Optics to record EL spectra. All EL spectra were measured with 1s integration time over 3 averages. The optical images and videos were taken and acquired by Thorlabs DCC1545M High Resolution USB2.0 CMOS Camera with weak illumination to enhance the optical contrast.

*Perovskite deposition:* The organic precursor methylammonium iodide $CH_3NH_3I$ was synthetized by mixing 10 ml of methylamine solution ($CH_3NH_2$, 40% in methanol, TCI) and 14 ml of hydroiodic acid (57% wt in water, Sigma-Aldrich). The reaction was accomplished in ice bath for 2 hours under magnetic stirring, and the solvent removed with a rotary evaporator (1 h at 60 mbar and 60 °C). The product was purified by dissolution in ethanol and recrystallization with diethylether, repeating the washing cycle 6 times. After filtration, the resulting white powder was dried in vacuum oven at 60 °C for 24 hours. Thin film of $CH_3NH_3PbI_3$ deposited on clean electrodes pre-patterned $SiO_2$ substrates. A 20% wt $CH_3NH_3PbI_3$ solution was prepared mixing stoichiometric amounts of $CH_3NH_3I$ and $PbI_2$ (99%, Sigma-Aldrich) in a solvent mixture of γ-butyrolactone and dimethylsulfoxide (7:3 volume ratio) and stirred overnight at 100 °C. In order to obtain continuous and uniform films, the solvent engineering technique was used.[14] The solution was spin-coated on the substrate using a 2 steps ramp: 1000 rpm for 10 s, 5000 rpm for 20 s. Toluene was drop-casted on the substrate during the second step. The resulting film was finally annealed at 100 °C for 30 minutes.

*Perovskite characterization:* Morphological analysis was performed through a FEI Helios 650 Nanolab Scanning electron microscope with 10 KV acceleration voltage. The X-Ray Diffraction (XRD) structural spectra were obtained using a diffractometer BRUKER D8 ADVANCE with Bragg-Brentano geometry employing Cu Kα radiation (l=1.54056 Å), a step increment of 0.02°, 1s of acquisition time and sample rotation of 5 $min^{-1}$.

*Computational Method:* The Density Function Theory (DFT) calculations have been carried out by the Perdew-Burke-Ernzerhof (PBE) generalized gradient approximation (GGA) using PWSCF code implemented in the Quantum ESPRESSO package.[53] For the structural optimization and band structure calculations, ultrasoft pseudopotentials including scalar-relativistic or full-relativistic effect were used to describe electron-ion interactions with electronic orbitals of H ($1s^1$); O, N and C ($2s^2$, $2p^2$); I ($5s^2$, $5p^2$) and Pb ($5d^{10}$, $6s^2$, $6p^2$).[54] The plane wave energy cutoff of wave function (charge) was set to be 40 (300) Ry. The crystal cell parameters were a = b = 8.81 Å, and c = 12.99 Å for tetragonal phase (Pm3m space group); and a = 8.77 Å, b = 8.56 Å and c = 12.97 Å for the orthorhombic phase (PNMA space group) of bulk $CH_3NH_3PbI_3$. The Monkhort-Pack scheme k-meshes are 4 × 4 × 4 for these two phases. The crystal cell and atomic positions were optimized until forces on single atoms were smaller than 0.01 eV/Å. The molecular graphics viewer VESTA was used to plot molecular structures.

The effective masses for electron ($m_e^*$) and hole ($m_h^*$) were estimated by fitting of the dispersion relation of $m^* = \hbar^2 \left[\frac{\partial^2 \varepsilon(k)}{\partial k^2}\right]^{-1}$ from band structures in **Figure 3** along the directions Γ-X, Γ-Z and Γ-M for tetragonal phase and Γ-X and Γ-Z for orthorhombic phase together with average values in these different routes. The carrier lifetime was evaluated by the semi-classical Boltzmann transport theory.[44] The only contribution of acoustic phonons was considered in evaluating scattering lifetime, where the charge carrier density ($n$) and mobility ($\mu$) are approximated as[55, 56]

$$n = \frac{(2m^* k_B T)^{3/2}}{2\pi^2 \hbar^3} \; {}^0F_0^{3/2}; \quad \mu = \frac{2\pi \hbar^4 eB}{m_I^* (2m_b^* k_B T)^{3/2} \Xi^2} \frac{3 \, {}^0F_{-2}^1}{{}^0F_0^{3/2}};$$

$$\text{where } {}^nF_l^m = \int_0^\infty \left(-\frac{\partial f}{\partial \zeta}\right) \zeta^n (\zeta + \alpha \zeta^2)^m [(1 + 2\alpha\zeta)^2 + 2]^{l/2} d\zeta;$$

$$f = 1/(e^{\zeta - \xi} + 1); \; \alpha = k_B T / E_g$$

$k_B$ is the Boltzmann constant, $e$ is the elementary charge, $T$ is the temperature, $\hbar$ is the Planck constant, and ξ is the reduced chemical potential; $m^*$ is the density of state effective mass, $m_I^*$ is the conductivity effective mass, $m_b^*$ is the band effective mass; $B$ is the bulk modulus ($B = \partial^2 E / \partial V^2$), $\Xi_{e-p/h-p}$ is the electron−phonon (or hole−phonon) coupling energy ($\Xi_{e-p/h-p} = V_0 (\Delta E_{CBM/VBM}/\Delta V)$, $n$, $m$, and $l$ power integer indices, $E_g$ is the electronic band gap, and ζ the reduced carrier energy.

**Acknowledgments**

Research was supported by NTU (NAP startup grant M4080511), the Singapore Ministry of Education (MOE2013-T2-044 and MOE2011-T3-1-005), and the Singapore-Berkeley Research Initiative for Sustainable Energy (SinBeRISE) CREATE Programme. The authors are grateful to Nripan Mathews, Pablo Boix, Mario Caironi and Annamaria Petrozza for the useful discussions, to Stefano Vezzoli and Saleem Umar for their help with electroluminescence measurements, and to Liu Hailong for assistance with SEM imaging.

**References**

1. Green MA, Ho-Baillie A, Snaith HJ. The emergence of perovskite solar cells. *Nat Photon* 2014, **8**(7)**:** 506-514.

2. Gao P, Gratzel M, Nazeeruddin MK. Organohalide lead perovskites for photovoltaic applications. *Energy & Environmental Science* 2014, **7**(8)**:** 2448-2463.

3. Lee M, Miyasaka T, Murakami T, Snaith H, Teuscher Jl. Efficient Hybrid Solar Cells Based on Meso-Superstructured Organometal Halide Perovskites. *Science* 2012, **338**(6107)**:** 643-647.

4. Stranks SD, Eperon GE, Grancini G, Menelaou C, Alcocer MJP, Leijtens T*, et al.* Electron-Hole Diffusion Lengths Exceeding 1 Micrometer in an Organometal Trihalide Perovskite Absorber. *Science* 2013, **342**(6156)**:** 341-344.

5. Xing G, Mathews N, Sun S, Lim S, Lam Y, Grätzel M*, et al.* Long-Range Balanced Electron- and Hole-Transport Lengths in Organic-Inorganic CH3NH3PbI3. *Science* 2013, **342**(6156)**:** 344-347.

6. D'Innocenzo V, Srimath Kandada AR, De Bastiani M, Gandini M, Petrozza A. Tuning the Light Emission Properties by Band Gap Engineering in Hybrid Lead Halide Perovskite. *Journal of the American Chemical Society* 2014, **136**(51)**:** 17730-17733.

7. Gonzalez-Pedro V, Juarez-Perez EJ, Arsyad W-S, Barea EM, Fabregat-Santiago F, Mora-Sero I*, et al.* General Working Principles of CH3NH3PbX3 Perovskite Solar Cells. *Nano Letters* 2014, **14**(2)**:** 888-893.

8. Liu M, Johnston MB, Snaith HJ. Efficient planar heterojunction perovskite solar cells by vapour deposition. *Nature* 2013, **501**(7467)**:** 395-398.

9. Zhou H, Chen Q, Li G, Luo S, Song T-b, Duan H-S*, et al.* Interface engineering of highly efficient perovskite solar cells. *Science* 2014, **345**(6196)**:** 542-546.

10. Kim H-S, Im SH, Park N-G. Organolead Halide Perovskite: New Horizons in Solar Cell Research. *The Journal of Physical Chemistry C* 2014, **118**(11)**:** 5615-5625.

11. Kojima A, Teshima K, Shirai Y, Miyasaka T. Organometal Halide Perovskites as Visible-Light Sensitizers for Photovoltaic Cells. *Journal of the American Chemical Society* 2009, **131**(17)**:** 6050-6051.

12. Burschka J, Pellet N, Moon S-J, Humphry Baker R, Gao P, Nazeeruddin M*, et al.* Sequential deposition as a route to high-performance perovskite-sensitized solar cells. *Nature* 2013, **499**(7458)**:** 316-319.

13. Boix PP, Nonomura K, Mathews N, Mhaisalkar SG. Current progress and future perspectives for organic/inorganic perovskite solar cells. *Materials Today* 2014, **17**(1)**:** 16-23.

14. Jeon NJ, Noh JH, Kim YC, Yang WS, Ryu S, Seok SI. Solvent engineering for high-performance inorganic–organic hybrid perovskite solar cells. *Nat Mater* 2014, **13**(9)**:** 897-903.

15. Heo JH, Im SH, Noh JH, Mandal TN, Lim C-S, Chang JA*, et al.* Efficient inorganic-organic hybrid heterojunction solar cells containing perovskite compound and polymeric hole conductors. *Nat Photon* 2013, **7**(6)**:** 486-491.

16. Salim T, Sun S, Abe Y, Krishna A, Grimsdale AC, Lam Y-M. Perovskite-Based Solar Cells: Impact of Morphology and Device Architecture on Device Performance. *Journal of Materials Chemistry A* 2014.

17. Luo J, Im J-H, Mayer MT, Schreier M, Nazeeruddin MK, Park N-G*, et al.* Water photolysis at 12.3% efficiency via perovskite photovoltaics and Earth-abundant catalysts. *Science* 2014, **345**(6204)**:** 1593-1596.

18. Tan Z-K, Moghaddam RS, Lai ML, Docampo P, Higler R, Deschler F*, et al.* Bright light-emitting diodes based on organometal halide perovskite. *Nat Nano* 2014, **9**(9)**:** 687-692.

19. Gil-Escrig L, Longo G, Pertegas A, Roldan-Carmona C, Soriano A, Sessolo M*, et al.* Efficient photovoltaic and electroluminescent perovskite devices. *Chemical Communications* 2015, **51**(3)**:** 569-571.

20. Hoye RLZ, Chua MR, Musselman KP, Li G, Lai M-L, Tan Z-K*, et al.* Enhanced Performance in Fluorene-Free Organometal Halide Perovskite Light-Emitting Diodes using Tunable, Low Electron Affinity Oxide Electron Injectors. *Advanced Materials* 2015**:** n/a-n/a.

21. Sutherland BR, Hoogland S, Adachi MM, Wong CTO, Sargent EH. Conformal Organohalide Perovskites Enable Lasing on Spherical Resonators. *ACS Nano* 2014, **8**(10)**:** 10947-10952.

22. Deschler F, Price M, Pathak S, Klintberg LE, Jarausch D-D, Higler R*, et al.* High Photoluminescence Efficiency and Optically Pumped Lasing in Solution-Processed Mixed Halide Perovskite Semiconductors. *The Journal of Physical Chemistry Letters* 2014, **5**(8)**:** 1421-1426.

23. Xing G, Mathews N, Lim SS, Yantara N, Liu X, Sabba D*, et al.* Low-temperature solution-processed wavelength-tunable perovskites for lasing. *Nat Mater* 2014, **13**(5)**:** 476-480.

24. Edri E, Kirmayer S, Mukhopadhyay S, Gartsman K, Hodes G, Cahen D. Elucidating the charge carrier separation and working mechanism of CH3NH3PbI3−xClx perovskite solar cells. *Nat Commun* 2014, **5**.

25. Bergmann VW, Weber SAL, Javier Ramos F, Nazeeruddin MK, Grätzel M, Li D, *et al.* Real-space observation of unbalanced charge distribution inside a perovskite-sensitized solar cell. *Nat Commun* 2014, **5**.

26. Giorgi G, Fujisawa J-I, Segawa H, Yamashita K. Small Photocarrier Effective Masses Featuring Ambipolar Transport in Methylammonium Lead Iodide Perovskite: A Density Functional Analysis. *The Journal of Physical Chemistry Letters* 2013, **4**(24)**:** 4213-4216.

27. Wehrenfennig C, Eperon GE, Johnston MB, Snaith HJ, Herz LM. High Charge Carrier Mobilities and Lifetimes in Organolead Trihalide Perovskites. *Advanced Materials* 2014, **26**(10)**:** 1584-1589.

28. Stoumpos CC, Malliakas CD, Kanatzidis MG. Semiconducting Tin and Lead Iodide Perovskites with Organic Cations: Phase Transitions, High Mobilities, and Near-Infrared Photoluminescent Properties. *Inorganic Chemistry* 2013, **52**(15)**:** 9019-9038.

29. Xiao Z, Yuan Y, Shao Y, Wang Q, Dong Q, Bi C, *et al.* Giant switchable photovoltaic effect in organometal trihalide perovskite devices. *Nat Mater* 2014, **advance online publication**.

30. Frost JM, Butler KT, Walsh A. Molecular ferroelectric contributions to anomalous hysteresis in hybrid perovskite solar cells. *APL Materials* 2014, **2**(8)**:** -.

31. Snaith HJ, Abate A, Ball JM, Eperon GE, Leijtens T, Noel NK, *et al.* Anomalous Hysteresis in Perovskite Solar Cells. *The Journal of Physical Chemistry Letters* 2014, **5**(9)**:** 1511-1515.

32. Shao Y, Xiao Z, Bi C, Yuan Y, Huang J. Origin and elimination of photocurrent hysteresis by fullerene passivation in CH3NH3PbI3 planar heterojunction solar cells. *Nat Commun* 2014, **5**.

33. Unger EL, Hoke ET, Bailie CD, Nguyen WH, Bowring AR, Heumuller T, *et al.* Hysteresis and transient behavior in current-voltage measurements of hybrid-perovskite absorber solar cells. *Energy & Environmental Science* 2014, **7**(11)**:** 3690-3698.

34. Kagan CR, Mitzi DB, Dimitrakopoulos CD. Organic-Inorganic Hybrid Materials as Semiconducting Channels in Thin-Film Field-Effect Transistors. *Science* 1999, **286**(5441)**:** 945-947.

35. Mitzi DB, Dimitrakopoulos CD, Kosbar LL. Structurally Tailored Organic−Inorganic Perovskites: Optical Properties and Solution-Processed Channel Materials for Thin-Film Transistors. *Chemistry of Materials* 2001, **13**(10)**:** 3728-3740.

36. Mitzi DB, Dimitrakopoulos CD, Rosner J, Medeiros DR, Xu Z, Noyan C. Hybrid field-effect transistor based on a low-temperature melt-processed channel layer. *Advanced Materials* 2002, **14**(23)**:** 1772-1776.

37. Noh JH, Im SH, Heo JH, Mandal TN, Seok SI. Chemical Management for Colorful, Efficient, and Stable Inorganic–Organic Hybrid Nanostructured Solar Cells. *Nano Letters* 2013, **13**(4)**:** 1764-1769.

38. Capelli R, Toffanin S, Generali G, Usta H, Facchetti A, Muccini M. Organic light-emitting transistors with an efficiency that outperforms the equivalent light-emitting diodes. *Nat Mater* 2010, **9**(6)**:** 496-503.

39. Verlaak S, Cheyns D, Debucquoy M, Arkhipov V, Heremans P. Numerical simulation of tetracene light-emitting transistors: A detailed balance of exciton processes. *Appl Phys Lett* 2004, **85**(12)**:** 2405-2407.

40. Kutes Y, Ye LH, Zhou YY, Pang SP, Huey BD, Padture NP. Direct Observation of Ferroelectric Domains in Solution-Processed $CH_3NH_3PbI_3$ Perovskite Thin Films. *J Phys Chem Lett* 2014, **5**(19)**:** 3335-3339.

41. Wang Q, Shao YC, Xie HP, Lyu L, Liu XL, Gao YL*, et al.* Qualifying composition dependent p and n self-doping in $CH_3NH_3PbI_3$. *Appl Phys Lett* 2014, **105**(16).

42. Zaumseil J, Sirringhaus H. Electron and ambipolar transport in organic field-effect transistors. *Chem Rev* 2007, **107**(4)**:** 1296-1323.

43. Quarti C, Grancini G, Mosconi E, Bruno P, Ball JM, Lee MM*, et al.* The Raman Spectrum of the $CH(3)NH(3)PbI(3)$ Hybrid Perovskite: Interplay of Theory and Experiment. *J Phys Chem Lett* 2014, **5**(2)**:** 279-284.

44. Wang H, Pei YZ, LaLonde AD, Snyder GJ. Weak electron-phonon coupling contributing to high thermoelectric performance in n-type PbSe. *Proceedings of the National Academy of Sciences of the United States of America* 2012, **109**(25)**:** 9705-9709.

45. Ma J, Wang LW. Nanoscale Charge Localization Induced by Random Orientations of Organic Molecules in Hybrid Perovskite $CH_3NH_3PbI_3$. *Nano Letters* 2014.

46. Frost JM, Butler KT, Brivio F, Hendon CH, van Schilfgaarde M, Walsh A. Atomistic Origins of High-Performance in Hybrid Halide Perovskite Solar Cells. *Nano Letters* 2014, **14**(5)**:** 2584-2590.

47. Swensen JS, Soci C, Heeger AJ. Light emission from an ambipolar semiconducting polymer field-effect transistor. *Appl Phys Lett* 2005, **87**(25)**:** 253511.

48. Muccini M. A bright future for organic field-effect transistors. *Nature Materials* 2006, **5**(8)**:** 605-613.

49. Chen ZY, Bird M, Lemaur V, Radtke G, Cornil J, Heeney M, *et al.* Origin of the different transport properties of electron and hole polarons in an ambipolar polyselenophene-based conjugated polymer. *Phys Rev B* 2011, **84**(11).

50. Wu KW, Bera A, Ma C, Du YM, Yang Y, Li L, *et al.* Temperature-dependent excitonic photoluminescence of hybrid organometal halide perovskite films. *Phys Chem Chem Phys* 2014, **16**(41): 22476-22481.

51. Even J, Pedesseau L, Katan C. Analysis of Multivalley and Multibandgap Absorption and Enhancement of Free Carriers Related to Exciton Screening in Hybrid Perovskites. *J Phys Chem C* 2014, **118**(22): 11566-11572.

52. Wang Y, Gould T, Dobson JF, Zhang HM, Yang HG, Yao XD, *et al.* Density functional theory analysis of structural and electronic properties of orthorhombic perovskite CH3NH3PbI3. *Phys Chem Chem Phys* 2014, **16**(4): 1424-1429.

53. Giannozzi P, Baroni S, Bonini N, Calandra M, Car R, Cavazzoni C, *et al.* QUANTUM ESPRESSO: a modular and open-source software project for quantum simulations of materials. *J Phys-Condens Mat* 2009, **21**(39).

54. Garrity KF, Bennett JW, Rabe KM, Vanderbilt D. Pseudopotentials for high-throughput DFT calculations. *Comp Mater Sci* 2014, **81**: 446-452.

55. Ravich YI, Efimova BA, Tamarche.Vi. Scattering of Current Carriers and Transport Phenomena in Lead Chacogenides. *Physica Status Solidi B-Basic Research* 1971, **43**(1): 11-&.

56. He YP, Galli G. Perovskites for Solar Thermoelectric Applications: A First Principle Study of CH3NH3AI3 (A = Pb and Sn). *Chemistry of Materials* 2014, **26**(18): 5394-5400.

# Supplementary Information

# Lead Iodide Perovskite Light-Emitting Field-Effect Transistor

Xin Yu Chin,[1] Daniele Cortecchia,[2,3] Jun Yin,[1,4] Annalisa Bruno,[1,3]
and Cesare Soci[1,4,*]

[1] *Division of Physics and Applied Physics, School of Physical and Mathematical Sciences, Nanyang Technological University, 21 Nanyang Link, Singapore 637371*

[2] *Interdisciplinary Graduate School, Nanyang Technological University, Singapore 639798*

[3] *Energy Research Institute @ NTU (ERI@N), Research Technoplaza, Nanyang Technological University, 50 Nanyang Drive, Singapore 637553*

[4] *Centre for Disruptive Photonic Technologies, Nanyang Technological University, Nanyang, 21 Nanyang Link, Singapore 637371*

*Corresponding author: csoci@ntu.edu.sg

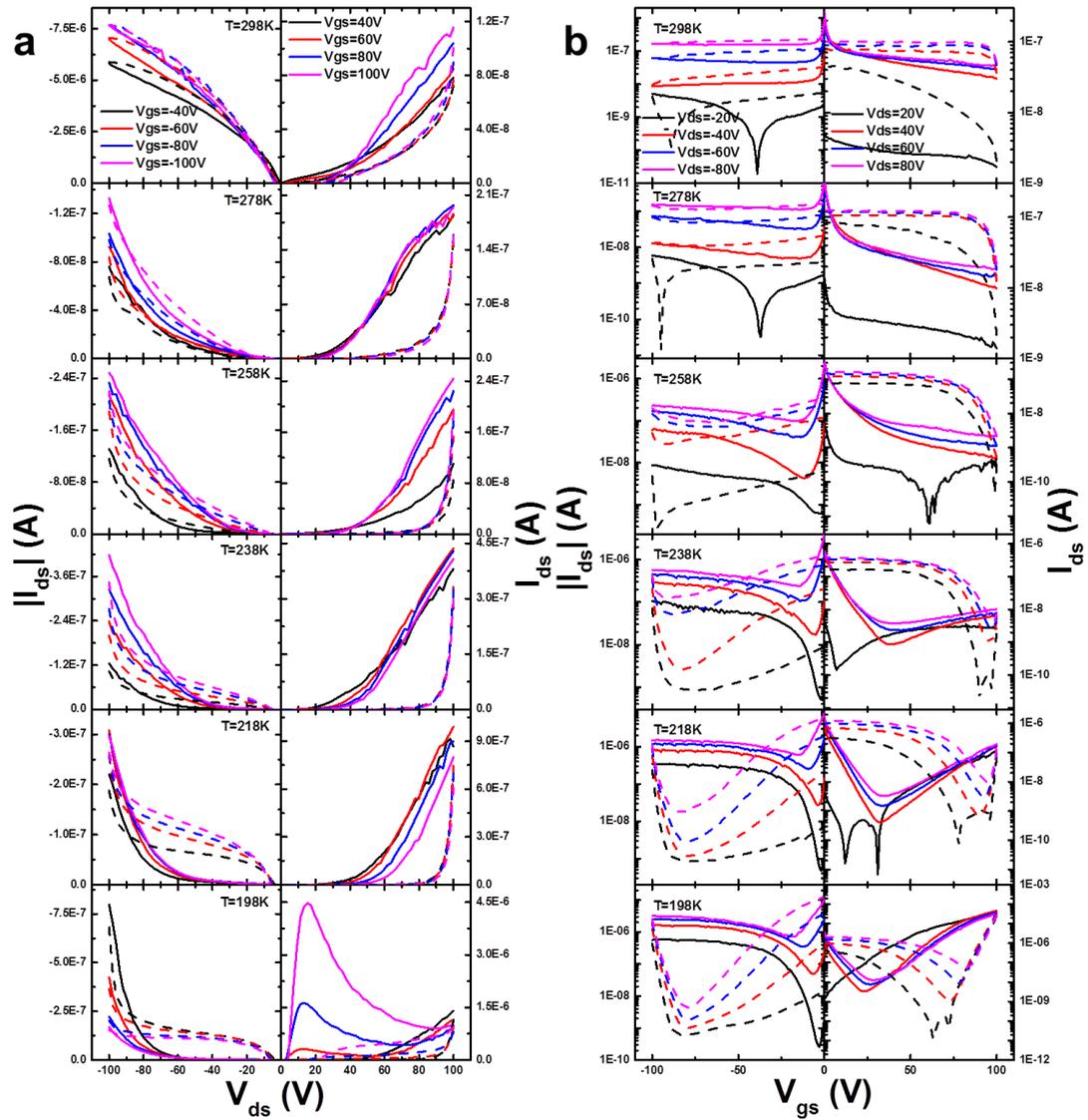

**Figure S1 | FET characteristics at 298 K, 278 K, 258 K, 238 K, 218 K, and 198 K. a**, FET output characteristics. The n-type output characteristics have been measured at $V_{gs}$ = 40 V to 100 V ($V_{gs}$ = 40 V black, $V_{gs}$ = 60 V red, $V_{gs}$ = 80 V blue, $V_{gs}$ = 100 V magenta), while the p-type output characteristics (left column) are measured at $V_{gs}$ = - 40 V to - 100 V ($V_{gs}$ = - 40 V black, $V_{gs}$ = - 60 V red, $V_{gs}$ = - 80 V blue, $V_{gs}$ = - 100 V magenta). **b**, FET transfer characteristics (ambipolar). The n-type transfer characteristics are measured at $V_{ds}$ = 20 V to 80 V ($V_{ds}$ = 20 V black, $V_{ds}$ = 40 V red, $V_{ds}$ = 60 V blue, $V_{ds}$ = 80 V magenta), while the p-type transfer characteristics (left column) are measured at $V_{ds}$ = - 20 V to – 80 V ($V_{ds}$ = - 20 V black, $V_{ds}$ = - 40 V red, $V_{ds}$ = - 60 V blue, $V_{ds}$ = - 80 V magenta). Solid and dashed curves are measured with forward and backward sweeping, respectively.

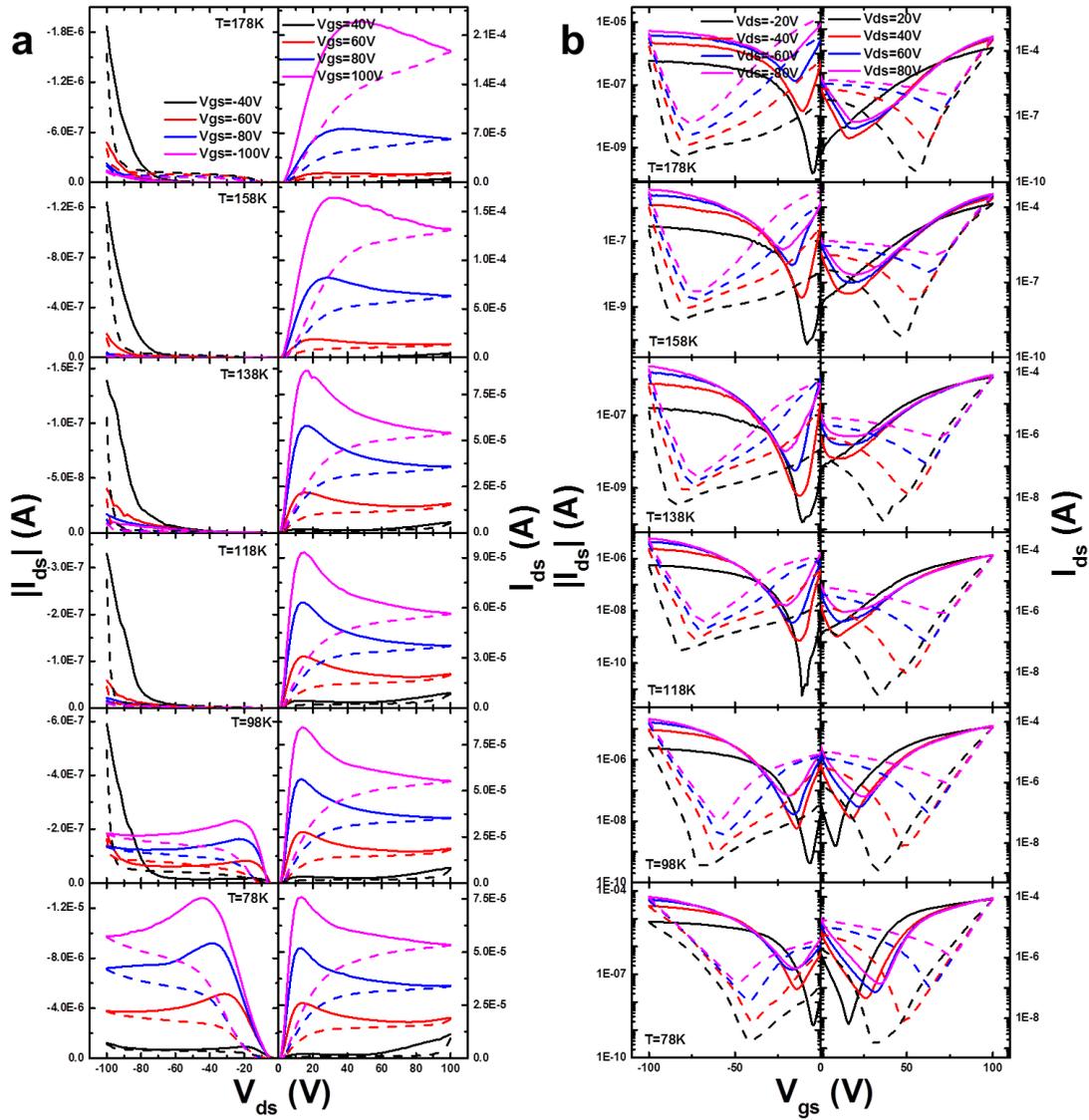

**Figure S2 | FET characteristics at 178 K, 158 K, 138 K, 118 K , 98 K, and 78 K. a**, FET output characteristics. The n-type output characteristics have been measured at $V_{gs}$ = 40 V to 100 V ($V_{gs}$ = 40 V black, $V_{gs}$ = 60 V red, $V_{gs}$ = 80 V blue, $V_{gs}$ = 100 V magenta), while the p-type output characteristics (left column) are measured at $V_{gs}$ = - 40 V to - 100 V ($V_{gs}$ = - 40 V black, $V_{gs}$ = - 60 V red, $V_{gs}$ = - 80 V blue, $V_{gs}$ = - 100 V magenta). **b**, FET transfer characteristics (ambipolar). The n-type transfer characteristics are measured at $V_{ds}$ = 20 V to 80 V ($V_{ds}$ = 20 V black, $V_{ds}$ = 40 V red, $V_{ds}$ = 60 V blue, $V_{ds}$ = 80 V magenta), while the p-type transfer characteristics (left column) are measured at $V_{ds}$ = - 20 V to – 80 V ($V_{ds}$ = - 20 V black, $V_{ds}$ = - 40 V red, $V_{ds}$ = - 60 V blue, $V_{ds}$ = - 80 V magenta). Solid and dashed curves are measured with forward and backward sweeping, respectively.

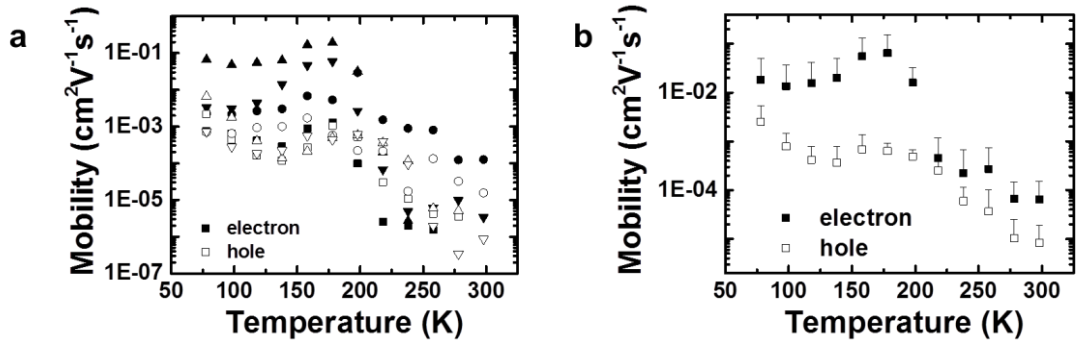

**Figure S3| Field effect mobilities across 4 different devices. a,** Field effect mobilities from 4 different devices, represented by square, circle, up triangle, down triangle. The filled symbols are electron mobilities, while the empty symbols are hole mobilities. **b,** Average mobilities and error bars obtained by averaging across 4 devices.

**Table S1| Calculated effective masses of charge carriers.** Estimated effective mass for electron and hole of $CH_3NH_3PbI_3$ from band structure including spin-orbital coupling effect.

| Phase | | $m_e^*$ | $m_h^*$ | Reduced Masses |
|---|---|---|---|---|
| Tetragonal | Γ-X | 0.178 | 0.261 | 0.106 |
| | Γ-Z | 0.284 | 0.474 | 0.177 |
| | Γ-M | 0.129 | 0.284 | 0.089 |
| | Average | 0.197 | 0.340 | 0.124 |
| Orthorhombic | Γ-X | 0.289 | 0.344 | 0.157 |
| | Γ-Z | 0.189 | 0.370 | 0.125 |
| | Average | 0.239 | 0.357 | 0.143 |

**Table S2| Required parameters for calculating moblities.** Band ($m_b^*$), conductivity ($m_I^*$) and density of state ($m^*$) effective mass, electron (hole)-phonon coupling (Ξ), and bulk modulus (*B*).

| | Tetragonal | | Orthorhombic | |
|---|---|---|---|---|
| | Electron | Hole | Electron | Hole |
| $m_b^*$ | 0.197 | 0.340 | 0.239 | 0.357 |
| $m_I^*$ | 0.157 | 0.290 | 0.163 | 0.288 |
| $m^*$ | 0.166 | 0.304 | 0.173 | 0.291 |
| Ξ (eV) | 7.2 | 8.4 | 6.8 | 7.4 |
| *B* (GPa) | 2.6 | 2.6 | 3.3 | 3.3 |

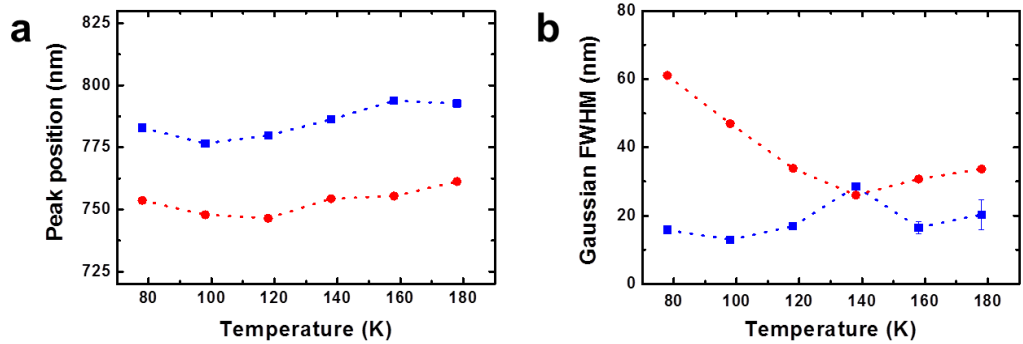

**Figure S4| Electroluminescence fitting parameters. a, b** Peak position (**a**) and FWHM (**b**) of Peak 1 (blue triangles) and Peak 2 (red circles) as a function of investigated temperature. The values are obtained by by fitting a deconvoluted double peak Gaussian function on **Figure 4**.